\documentclass{amsart}

\usepackage{tikz}

\theoremstyle{definition}

\theoremstyle{remark}

\numberwithin{equation}{section}

\begin{document}

\title{Subfactors and Mathematical Physics}

%    Information for first author
\author{David E Evans}
%    Address of record for the research reported here
\address{School of Mathematics, Cardiff University, Senghennydd Road, Cardiff CF24 4AG, Wales, United Kingdom}
%    Current address
%\curraddr{Department of Mathematics and Statistics,
%Case Western Reserve University, Cleveland, Ohio 43403}
\email{EvansDE@cardiff.ac.uk}
%    \thanks will become a 1st page footnote.
%\thanks{The first author was supported in part by NSF Grant \#000000.}

%    Information for second author
\author{Yasuyuki Kawahigashi}
\address{Department of Mathematical Sciences,
The University of Tokyo,
3-8-1 Komaba, Tokyo, 153-8914,
Japan}
\email{yasuyuki@ms.u-tokyo.ac.jp}
\thanks{The second author was partially supported by 
JST CREST program JPMJCR18T6 and
Grants-in-Aid for Scientific Research 19H00640
and 19K21832.}

%    General info

\date{xxx xxx, 2022}

\dedicatory{This paper is dedicated to the memory of Vaughan Jones}


\begin{abstract}
This paper  surveys the  long-standing connections and impact  between Vaughan Jones's theory of
subfactors and various topics in mathematical physics, namely 
statistical mechanics, quantum field theory, quantum 
information and two-dimensional conformal field theory.
\end{abstract}

\maketitle

\section{Subfactors and mathematical physics}

Subfactor theory was initiated by  Vaughan Jones \cite{J1}. This led him
to the study of a new type of \textsl{quantum symmetry}.  This notion
of quantum symmetries led to  a diverse range of applications
including the Jones polynomial, a completely new
invariant in knot theory which led to the new field of  {\it quantum topology}.  His novel theory
has deep connections to various topics in mathematical physics.
This  renewed interest in known connections between
mathematical physics and operator algebras, and opened up totally novel
frontiers.  
We present a survey on these interconnecting
topics with emphasis on  statistical mechanics and
quantum field theory, particularly two-dimensional conformal 
field theory.

\section{Subfactors and statistical mechanics}

Let $N\subset M$ be a subfactor of type II$_1$ and $[M:N]$
its \textsl{Jones index}, which is a positive real number or infinity.
That is, $N$ and $M$ are infinite-dimensional simple von Neumann algebras 
with a trace $\mathrm{tr}$.
We only consider the case that $[M:N]$ is finite.
Vaughan  \cite{J1} constructed a sequence of projections $e_j$,
$j=1,2,3,\dots$, called the \textsl{Jones projections}, and
discovered the following relations:
\begin{equation}
\label{TLJ}
\left\{ \,
\begin{aligned}
e_j&=e_j^2=e_j^*,\\
e_j e_k&=e_k e_j,\quad j\neq k,\\
e_j e_{j\pm1} e_j&=[M:N]^{-1}e_j.
\end{aligned}
\right.
\end{equation}
Using these relations and a trace, Vaughan showed that the set
of possible values of the Jones indices is exactly equal to
\[
\left\{4\cos^2\frac{\pi}{n}\mid n=3,4,5,\dots\right\}\cup [4,\infty].
\]

  Vaughan made the substitution 
\begin{equation}
\label{braid}
 \sigma_j = t e_j-(1-e_j)
\end{equation}
where $[M:N] = \lambda^{-1} = 2 +t + t^{-1}$ to yield the  Artin 
relations of the braid group, where $\sigma_j$ is the braid which interchanges the $j$ and $j+1$ strands. 
Vaughan's representation came equipped with a trace $\mathrm{tr}$  satisfying the
Markov trace property in the probabilistic sense $\mathrm{tr}(xe_j)=[M:N]^{-1}\mathrm{tr}(x)$, where
$x$ belongs to the algebra generated by $e_1,e_2,\dots,e_{j-1}$.
Any link arises as a closure of a braid by a theorem of Alexander,
and two braids give the
same link if and only if they are related by a series of 
two types of moves, known as the \textsl{Markov moves}, by a theorem
of Markov.
The trace property $\mathrm{tr}(xy)=\mathrm{tr}(yx)$ and the
above Markov trace property give invariance of a certain adjusted
trace value of a braid under the two Markov moves.  This is
the Jones polynomial \cite{J2, J3} in the variable $t$  a polynomial invariant of a link.

Evans pointed out in 1983 that these relations (\ref{TLJ}) appear in similar formalism to 
one studied by Temperley-Lieb \cite{TL} in solvable statistical mechanics. The \textsl{Yang-Baxter equation} plays an important role in subfactor theory and quantum groups.
The two-dimensional Ising model assigns two possible spin values $\pm$ at the vertices of a lattice. Important generalisations include the Potts model, with $Q$ states at each vertex, and vertex  or IRF (interaction round a face) models, where the degrees of freedom are assigned to the edges of the lattice.  The transfer matrix method, originated by Kramers and Wannier, assigns  a matrix of Boltzmann weights  to a one dimensional row
 lattice. The partition function of a rectangular lattice in general is then obtained by gluing together matrix products
of the transfer matrix.  Baxter \cite{Ba} showed how to construct commuting families of transfer matrices via Boltzmann weights satisfying the Yang-Baxter Equation YBE.  The YBE is an enhancement of the braid relations in (\ref{braid}), as it reduces to them in a certain limit. This commutativity permits simultaneous  diagonalisation, with the largest eigenvalue being crucial for computing the free energy.  The transfer matrix method transforms the classical statistical mechanical model to a one-dimensional quantum model.
A conformal field theory can arise from the scaling limit of a statistical mechanical lattice model at criticality.
 Temperley and Lieb \cite{TL} found that the transfer matrices of the Potts model and an ice-type vertex model could both  be described through generators obeying the same relations as in Vaughan's work (\ref{TLJ})
 and  in this way demonstrated   equivalence of the models.
Whilst the relations for the Potts model only occur when $\lambda^{-1} =Q$ is  integral, the partition function is a Tutte-Whitney  dichromatic polynomial. One variable is $Q$ which can be extrapolated and the partition function is then related to the Jones polynomial on certain links  associated to the lattice \cite[page 108]{J2}. The ice-type representation though has  a continuous parameter $\lambda$. In both cases, the Markov trace did not manifest itself.

Pimsner and Popa \cite{PP} discovered that the inverse of the index  namely $[M:N]^{-1}$ is the best constant $c\geq 0$ for which $E_N(x^*x) \geq c x^*x,$ for all $ x\in M$, which they  called the {\it probabilistic index}. Here $E_N$ is the conditional expectation of $M$ onto $N$ which gives rise to the first Jones projection in the tower.
This was key to creating the link with the theory of Doplicher-Haag-Roberts, by Longo \cite{L1} identifying statistical dimension with the Jones index, and by Fredenhagen, Rehren and Schroer \cite{FRS1, FRS2}, in the late 80's, and also key to calculating all of the subsequent entropy quantities/invariants related to subfactors, including the calculation of the entropy of the shift on the Jones projections  and the calculation  of the Connes-Stormer entropy \cite{CSt}, $H(M|N)=\ln([M:N])$, for irreducible subfactors.

 For a subfactor $N\subset M$ with finite Jones index, we have the
\textsl{Jones tower construction}
\[
N\subset M\subset M_1\subset M_2\subset \cdots,
\]
where $M_k$ is generated by $M$ and $e_1,e_2,\dots,e_k$. The basic construction from $N \subset M$ to $M \subset M_1$ and its iteration to give the Jones tower of II$_1$ factors has a fundamental role in subfactor theory and applications in mathematical physics.
The \textsl{higher relative commutants} $M'_j\cap M_k$, $j\le k$, give a system of {\it commuting squares of inclusions} of 
finite dimensional $C^*$-algebras with a trace, an object denoted by $\mathcal G_{N\subset M}$ and called the \textsl{standard invariant} of $N\subset M$. 
This exceptionally rich mathematical structure encodes algebraic
and combinatorial information about the subfactor, a key component of which is  
a connected, possibly infinite bipartite graph $\Gamma_{N\subset M}$, of Cayley type, called the \textsl{principal graph} of $N\subset M$, 
with a canonical weight vector $\vec{v}$, whose entries are square roots of indices of irreducible inclusions in the Jones tower. 
The weighted graph 
$(\Gamma, \vec{v})$ satisfies 
the Perron-Frobenius type condition $\Gamma^t\Gamma(\vec{v})=[M:N]\vec{v}$, and  also $\|\Gamma\|^2\leq [M:N]$. 

Of particular relevance to mathematical physics is when $N\subset M$ has \textsl{finite depth}, corresponding to the graph $\Gamma$ being finite, 
in which case the weights $\vec{v}$ give the (unique) Perron-Frobenius eigenvector, entailing $\|\Gamma\|^2=[M:N]$. 
Finite depth is automatic when the index $[M:N]$ is less than $4$, 
where indeed all bipartite graphs are finite and have norms of the form $2\cos^2(\pi/n)$, $n\geq 3$.  

The objects $\mathcal G_{N\subset M}$ have been axiomatised in a number of ways, by Ocneanu with paragroups and 
connections \cite{O} in the finite depth case, then in the general case by Popa with $\lambda$-lattices \cite{P2} and by Vaughan with  planar algebras \cite{J12}. 

By Connes fundamental result in \cite{Co2}, the \textsl{hyperfinite} II$_1$ factor $R$, obtained as an inductive limit of finite dimensional algebras, 
is the unique amenable II$_1$ factor, so in particular all its finite index subfactors are isomorphic to $R$. In a series of papers \cite{P1, P3, P6, P7}, 
Popa identified the appropriate notion of amenability for inclusions of II$_1$ factors $N\subset M$ and for the objects $\mathcal G_{N\subset M}$, 
in several equivalent ways, one of which being the Kesten-type condition $\|\Gamma_{N\subset M}\|^2=[M:N]$. He proved the important result 
that  for hyperfinite subfactors $N\subset M$ satisfying this amenability condition, $\mathcal G_{N\subset M}$ is a complete invariant. In other words, whenever 
$M\simeq R$ and $\|\Gamma_{N\subset M}\|^2=[M:N]$ (in particular if $N\subset M$ has finite depth),  
$N\subset M$ can be recovered from the data encoded by the sequence of commuting squares in the Jones tower.

Constructions of interesting commuting squares are
related to statistical mechanics through the Yang-Baxter equation and
an \textsl{IRF} , vertex or spin
model \cite{J4}.  (See  the monograph of Baxter \cite{Ba} for this type of statistical
mechanical models.  Also see \cite{J7} for a
general overview   by Vaughan on this type of relations.)
 We choose one edge each from the
four diagrams for the four inclusions so that they make
a closed square.  Then we have an assignment of a complex
number to each such square.
Ocneanu \cite{O} gave a combinatorial characterisation of
this assignment of complex numbers under the name of a \textsl{paragroup}
and a \textsl{flat connection}.  We also assign a complex number,
called a \textsl{Boltzmann weight}, to each square arising from
a finite graph in the theory of IRF or vertex models and we have
much similarity between the two notions.
The simplest example  corresponds to the \textsl{Ising model} built on the Coxeter-Dynkin diagram $A_3$
and a more general case corresponds to the
Andrews-Baxter-Forrester model \cite{ABF}
related to the quantum groups $U_q(sl_2)$ for $q=\exp(2\pi i/l)$ a root of unity. 
These fundamental examples correspond to the subfactors
generated by the Jones projections alone and the graphs for 
these cases are the \textsl{Coxeter-Dynkin diagrams} of type $A_n$.
Others  related to the quantum groups $U_q(sl_n)$ have been studied in
\cite{JMO, DZ}.

We give a typical example of a flat connection as follows.
Fix one of the Coxeter-Dynkin diagrams of type $A_n$, $D_{2n}$, $E_6$
or $E_8$ and use it for the four diagrams.  
Let $h$ be its Coxeter number and set
$\varepsilon=\sqrt{-1}\exp(\pi\sqrt{-1}/2h)$.
We write $\mu_j$ for the Perron-Frobenius eigenvector
entry for a vertex $j$ for the adjacency of the diagram.
Then the flat connection is given as in Fig.\ref{Dynkin} and is essentially a normalisation of the  braid element (\ref{braid}):

\begin{figure}[h]
\begin{center}
\begin{tikzpicture}[scale=0.8]
\draw [thick, ->] (1,1)--(2,1);
\draw [thick, ->] (1,2)--(2,2);
\draw [thick, ->] (1,2)--(1,1);
\draw [thick, ->] (2,2)--(2,1);
\draw (1.5,1.5)node{$W$};
\draw (1,1)node[below left]{$l$};
\draw (1,2)node[above left]{$j$};
\draw (2,1)node[below right]{$m$};
\draw (2,2)node[above right]{$k$};
\draw (5.8,1.5)node{$\displaystyle=\delta_{kl}\varepsilon+
\sqrt{\frac{\mu_k \mu_l}{\mu_j\mu_m}}\delta_{jm}\bar\varepsilon$};
\end{tikzpicture}
\end{center}
\caption{A flat connection on the Coxeter-Dynkin diagram}
\label{Dynkin}
\end{figure}
The index value given by this construction is $4\cos^2 (\pi/h)$.
If the graph is $A_n$, then the vertices are labeled with
$j=1,2,\dots,n$ and the Perron-Frobenius eigenvector entry
for the vertex $j$ is given by $\sin (j\pi/(n+1))$.  The
value in Fig.\ref{Dynkin} in this case is essentially the 
same as what the Andrews-Baxter-Forrester model gives at a limiting value and
it also arises from a specialisation
of the \textsl{quantum $6j$-symbols} for $U_q(sl_2)$ at a root of unity
in the sense that two of the ``$6j$''s are chosen to be
the fundamental representation of $U_q(sl_2)$.
These are also related to IRF models   by Roche in \cite{R}.
These subfactors for the Dynkin diagrams $A_n$ are
the ones constructed by Vaughan \cite{J1} as
$N=\langle e_2, e_3,\dots \rangle$ and
$M=\langle e_1, e_2, e_3, \dots \rangle$ with the above 
relations (\ref{TLJ}) with $[M:N]=4\cos^2(\pi/(n+1))$.

The same formula as in Fig.\ref{Dynkin} for the Coxter-Dynkin
diagrams $D_{2n+1}$ and $E_7$ almost gives a flat connection,
but the flatness axiom fails.  There are corresponding
subfactors but they have principal graphs $A_{4n-1}$ and $D_{10}$ respectively.  Nevertheless, the diagrams $D_{2n+1}$ and $E_7$ have
 interesting interpretations in connection with non-local
extensions of \textsl{conformal nets} $SU(2)_k$, as explained below.

The relations (\ref{TLJ}) of the Jones projections $e_j$
are reminiscent of the defining relations
of the \textsl{Hecke algebra}
$H_n(q)$ of type $A$ with complex parameter 
$q$, which is the free complex algebra generated by $1,g_1,g_2,\dots,
g_{n-1}$ satisfying 
\begin{equation*}
\left\{ \,
\begin{aligned}
g_j g_{j+1} g_j&=g_{j+1} g_j g_{j+1},\\
g_j g_k &= g_k g_j,\quad\text{for\ $j\neq k$},\\
g_j^2&=(q-1)g_j+q.
\end{aligned}
\right.
\end{equation*}
This similarity was exploited to 
construct more examples of subfactors with index values
$\displaystyle\frac{\sin^2 (k\pi/l)}{\sin^2 (\pi/l)}$ with
$1\le k\le l-1$ in the early days of
subfactor theory by Wenzl in a   University of Pennsylvania  thesis supervised by Vaughan
\cite{We}.  He constructed representations $\rho$
of $H_\infty(q)=\bigcup_{n=1}^\infty H_n(q)$ with roots of unity  $q=\exp(2\pi i/l)$
and $l=4,5,\dots$ such that $\rho(H_n(q))$ is
always semi-simple and gave a subfactor as
$\rho(\langle g_2,g_3,\dots\rangle)''\subset 
\rho(\langle g_1,g_2,\dots\rangle)''$ using a suitable trace.
The index values converge
to $k^2$ as $l\to\infty$.  When $k=2$, these subfactors are the
ones constructed by Vaughan for the Coxeter-Dynkin diagram $A_{l-1}$.  This 
construction is also understood in the context of IRF 
models \cite{DZ, JMO} related to $SU(k)$.
The relation between the Hecke algebras and the quantum groups
$U_q(sl_n)$ is a ``quantum'' version of the classical \textsl{Weyl duality}.
This duality also connects this Jones-Wenzl approach based on
statistical mechanics and type II$_1$ factors with the Jones-Wassermann approach based
on quantum field theory and type III$_1$ factors which is explained below.

It is important to have a \textsl{spectral parameter} for the
Boltzmann weights satisfying the
Yang-Baxter equation in \textsl{solvable lattice models}, but we
do not have such a parameter for a flat connection
initially in subfactor theory.
We usually obtain a flat connection by a certain specialisation
of a spectral parameter for a Boltzmann weight.
Vaughan proposed ``Baxterization'' in \cite{J5} for the converse
direction in the sense of
introducing a parameter  for analogues of the
Boltzmann weights in subfactor theory.  This is an idea to
obtain a physical counterpart from a subfactor, and we discuss a 
similar approach to construct a  conformal field theory from a given
subfactor at the end of this article. It should be noted that to rigorously construct a 
conformal field theory at criticality is  a notoriously difficult problem -- even for the Ising model,
 see e.g.\ \cite{SMJ}.

The finite depth condition means that we have a finite graph
in this analogy to solvable lattice models.  Even from
a set of algebraic or combinatorial data similar to
integrable lattice models involving infinite graphs,
one sometimes constructs a corresponding subfactor. 
 A major breakthrough of  Popa \cite{P5}  was to show that the Temperley-Lieb-Jones lattice is indeed a standard invariant showing for the first time that for 
any index  greater than $4$ that  there exist subfactors with just the Jones projections as the higher relative commutants. 
Then, introducing tracial amalgamated free products, Popa \cite{P2} could show  existence in full generality. These papers \cite{P5, P2} led to important links with free probability theory, leading to more sophisticated free random models to prove that certain amalgamated free products are free group factors and
adapted, by Ueda \cite{U}, to prove similar existence/reconstruction statements for actions of quantum groups. Popa and Shlyakhtenko \cite{PSh} showed that  any $\lambda$-lattice acts on the free group factor $L(\mathbb F_\infty)$. This involved a new construction  of subfactors from $\lambda$-lattices, starting from a commuting square of semifinite von Neumann algebras, each one  a direct sum of type I$_\infty$ factors with a semifinite trace,  and with free probability techniques showing that the factors resulting from this construction are $\infty$-amplifications of $L(\mathbb F_\infty)$.
 The von Neumann algebras resulting in these
constructions are not hyperfinite.  A new proof using graphical tools, probabilistic methods and planar algebras was
later found by Guionnet-Jones-Shlyakhtenko \cite{GJS}.
Moreover  they and Zinn-Justin \cite{GJSZ} use matrix model computations in loop models of statistical
mechanics and graph planar algebras to construct novel matrix models for Potts models
on random graphs.
This is based on the \textsl{planar algebra} machinery 
developed by Vaughan \cite{J12} for understanding higher relative commutants
of subfactors.  In \cite{GJS2} Guionnet-Jones-Shlyakhtenko  explicitly show that it is the same construction as in the Popa-Shlyakhtenko \cite{PSh} paper.
The paper \cite{J12} has been published only very
recently in the Vaughan Jones memorial special issue after his passing away, 
but its preprint version appeared in 1999 and 
has been highly influential. 
Note also that Kauffman \cite{Kff, Kff2} had found a diagrammatic  construction of  the Jones polynomial directly related to the Potts model  based on  a diagrammatic presentation of the Temperley-Lieb algebra which then has a natural home in the planar algebra formalism. The polynomial was
understood by Reshetikhin-Turaev in \cite{RT} in the context of representations
of the quantum groups $U_q(sl_2)$  \cite{D, Ji}.

\section{Subfactors and quantum field theory}

Witten \cite{W} gave a new interpretation of the Jones polynomial
based on quantum field theory, the Chern-Simons gauge field theory,
and generalised it to an invariant
of a link in a compact 3-manifold. However,  it was not clear why
we should have a \textsl{polynomial} invariant in this way.  
Taking an empty link,
yields an invariant of a compact 3-manifold.  Witten used a path integral
formulation and was not mathematically rigorous.  A mathematically
well-defined version based on combinatorial arguments using 
\textsl{Dehn surgery} and the \textsl{Kirby calculus} has been given by
Reshetikhin and Turaev \cite{RT}.
In the case of an empty link, we realise a 3-manifold
from a \textsl{framed link} with the Dehn surgery, make a weighted sum
of invariants of this link using representations of
a certain quantum group
at a root of unity and prove that this weighted sum is invariant
under the Kirby moves.  Two framed links give homeomorphic
manifolds if and only if they are related with a series of
Kirby moves.
For the quantum group $U_q(sl_2)$,
the link invariant  is the \textsl{colored Jones polynomial}.
A \textsl{color} is a representation of the quantum group
and labels a connected component of a link.
This actually gives a $(2+1)$-dimensional
\textsl{topological quantum field theory} in the sense of Atiyah \cite{A}, which is a certain mathematical
axiomatisation of a quantum field theory based on topological
invariance.  Roughly speaking, we assign
a finite dimensional Hilbert space to each closed $2$-dimensional
manifold, and also assign a linear map from one such Hilbert
space to another to a cobordism so that this assignment is functorial.
It is also easy to extend this construction from quantum groups to
general \textsl{modular tensor categories} as we explain below.

A closely related, but  different, $(2+1)$-dimensional
topological quantum field theory has been given by 
Turaev and Viro \cite{TV}.  In this formulation, one triangulates
a 3-manifold, considers a weighted sum of quantum $6j$-symbols arising
from a quantum group depending on the triangulation, and
proves that this sum is invariant under the \textsl{Pachner moves}.
Two triangulated manifolds are homeomorphic to each other if and
only if we obtain one from the other with a series of Pachner moves.
This has been generalised to another $(2+1)$-dimensional topological
quantum field theory using \textsl{quantum $6j$-symbols} arising 
from a subfactor by Ocneanu.  (See \cite[Chapter 12]{EK}.)
Here we only  need a \textsl{fusion category} structure which
we explain below, and no \textsl{braiding}.
This is different from the above Reshetikhin-Turaev case.
For a given fusion category, we apply the Drinfel$'$d center
construction, a kind of  ``quantum double'' construction,
to get a modular tensor category with a non-degenerate braiding.  This construction was developed in subfactor theory  by
Ocneanu \cite{O} through an asymptotic inclusion, by Popa \cite{P3} through a symmetric enveloping algebra, through the Longo-Rehren subfactor \cite{LR} and 
Izumi  \cite{I1, I2} and in a categorical setting by M\"uger \cite{Mug}.  We  then
apply the Reshetikhin-Turaev construction to  the double.  We can also
apply the Turaev-Viro-Ocneanu construction to the   original fusion category,
and these two procedures give the same topological quantum field theory
\cite{KSW}.  In particular, if we start with $U_q(sl_2)$ at a
root of unity, the Turaev-Viro invariant of a closed $3$-manifold
is the square of the absolute value of the Reshetikhin-Turaev
invariant of the same $3$-manifold.

Another connection of subfactors to quantum field theory is
through \textsl{algebraic quantum field theory}, which is a bounded operator
algebraic formulation of quantum field theory.  
The usual ingredients for describing a quantum field theory are 
as follows.

\begin{enumerate}
\item A spacetime, such as the 4-dimensional \textsl{Minkowski space}.
\item A spacetime symmetry group, such as the \textsl{Poincar\'e group}.
\item A Hilbert space of states, including the \textsl{vacuum}.
\item A projective unitary representation of the
spacetime symmetry group on the Hilbert space of states.
\item A set of quantum fields, that is,
operator-valued distributions
defined on the spacetime acting on the Hilbert space of 
states.
\end{enumerate}

An ordinary distribution assigns a number to each test function.
An operator-valued distribution assigns a (possibly unbounded) operator
to each test function.
The \textsl{Wightman axioms} give a direct axiomatisation using these and
they have a long history of research, but it is technically
difficult to handle operator-valued distributions, so we have
a different approach based on bounded linear operators
giving observables.  Let $O$ be a region within the spacetime.
Take a quantum field $\varphi$ and a test function $f$ supported
on $O$.  The self-adjoint part of $\langle \varphi, f\rangle$ is 
an observable in $O$ which could be unbounded.  Let
$A(O)$ denote the von Neumann algebra generated by spectral projections
of such self-adjoint operators.  This passage from operator-valued
distributions to von Neumann algebras is also used in the construction
of a \textsl{conformal net} from a \textsl{vertex operator algebra}  by Carpi-Kawahigashi-Longo-Weiner
\cite{CKLW} which we explain below.  Note that a von Neumann algebra
contains only bounded operators.  

Locality is an important axiom
arising from the Einstein causality which says that if two
regions are spacelike separated, observables in these regions
have no interactions, hence the corresponding operators commute.
In terms of the von Neumann algebras $A(O)$, we require that
$[A(O_1), A(O_2)]=0$, if $O_1$ and $O_2$ are spacelike
separated, where the Lie bracket means the commutator.
This family of von Neumann 
algebras parameterised by spacetime regions is called a \textsl{net
of operator algebras}.  Algebraic quantum field theory gives an
axiomatisation of a net of operator algebras, together with
a projective unitary representation of a spacetime symmetry
group on the Hilbert space of states including the vacuum.
A main idea is that it is not each von Neumann algebra
but the relative relations among these von Neumann algebras that
contains the physical contents of a quantum field theory.
In the case of two-dimensional conformal field theory, which is
a particular example of a quantum field theory, each von Neumann
algebra $A(O)$ is always a hyperfinite type III$_1$ factor, which
is unique up to isomorphism and is the \textsl{Araki-Woods
factor} of type III$_1$ . Thus the isomorphism class of a single von Neumann algebra
contains no physical information.  Each local algebra of a conformal net is a factor of type III$_1$ by
\cite[Proposition 1.2]{GL}.  It is also hyperfinite because it has
a dense subalgebra given as an increasing union of type I algebras,
which follows from the split property shown in \cite[Theorem 5.4]{MTW}.

Fix a net $\{A(O)\}$ of von Neumann algebras.  It has a natural
notion of a representation on another Hilbert space without the
vacuum vector.  The action of these von Neumann algebras
on the original Hilbert space itself is a representation and it
is called the \textsl{vacuum representation}.
We also have natural notions of \textsl{unitary equivalence}
and \textsl{irreducibility} of 
representations.  The unitary equivalence
class of an irreducible representation of the net $\{A(O)\}$
is called a \textsl{superselection sector}.  We also have a direct
sum and irreducible decomposition for representations.
If we have two representations of a group, it is very easy
to define their tensor product representation, but it is not clear
at all how to define a tensor product representation of two
representations of a single net of operator algebras.
Doplicher-Haag-Roberts gave a proper definition of
the tensor product of two representations
\cite{DHR1, DHR2}.  Under a certain natural assumption, each
representation has a representative given by an endomorphism 
of a single algebra $A(O)$ acting on the vacuum Hilbert space
for some fixed $O$.  This endomorphism
contains complete information about the original representation.
For two such endomorphisms $\rho$ and $\sigma$, the composed
endomorphism $\rho\sigma$ also corresponds to a representation
of the net $\{A(O)\}$.  This gives a correct notion of the
tensor product of two representations. Furthermore, it turns out
that the two compositions
$\rho\sigma$ and $\sigma\rho$ of endomorphisms give unitarily equivalent
representations.  
If the spacetime dimension is higher than 2, this commutativity
of the tensor product is similar to unitary equivalence of
$\pi_1\otimes\pi_2$ and $\pi_2\otimes \pi_1$ for two
representations $\pi_1$ and $\pi_2$ of the same group.
The representations now
give a \textsl{symmetric monoidal $C^*$-category}, where
a representation gives an object, an intertwiner gives a
morphism, and the above composition of endomorphisms gives
the tensor product structure.
This category produces a compact group from
the new duality of Doplicher-Roberts \cite{DR}.  Here an object
of the category is an endomorphism and a morphism in
$\mathrm{Hom}(\rho,\sigma)$ is an intertwiner, that is, an
element in 
\[
\{T\in A(O)\mid T\rho(x)=\sigma(x)T\text{\ for\ all\ }
x\in A(O)\}.
\]
In other words, the 
Doplicher-Roberts duality gives an abstract characterisation of
the representation category of a compact group among general tensor
categories.  The vacuum representation plays the role of the
trivial representation of a group, and the \textsl{dual representation}
of a net of operator algebras corresponds to the dual representation
of a compact group.  This duality is related to the classical
\textsl{Tannaka duality}, but gives a duality  more generally for
abstract tensor categories.

Using the structure of a symmetric monoidal $C^*$-category, we
define a \textsl{statistical dimension} of each representation,
which turns out to be a positive integer or infinity \cite{DHR1, DHR2}.
That the Jones index value takes on only discrete values
below 4 is reminiscent of this fact that
a statistical dimension can take only integer values. 
Longo \cite{L1, L2} showed that the statistical
dimension of the representation corresponding to an endomorphism
$\rho$ of $A(O)$ is equal to the square root of the Jones
index $[A(O):\rho(A(O))]$.  This  opened up a wide range of new 
interactions between subfactor theory and algebraic quantum
field theory.  

Generalizing the notion of a superselection sector,
Longo \cite{L1, L2} introduced the notion of a \textsl{sector}, 
the unitary equivalence class of an endomorphism of a factor of type III, inspired by
 Connes theory of correspondences, based on the equivalences between Hilbert bimodules,  endomorphisms and positive definite functions on doubles  \cite{Co4} \cite[VB]{Co3}, \cite{CoJo} and  see e.g. Popa \cite{P8} for developments. 
He defined a \textsl{dual sector} using the 
\textsl{canonical endomorphism} which he had 
introduced based on the modular conjugation in
Tomita-Takesaki theory.  Note that in a typical situation of a subfactor
$N\subset M$, these von Neumann algebras are isomorphic, so we have
an endomorphism $\rho$ of $M$ onto $N$.  Then we have the
dual endomorphism $\bar\rho$, and the irreducible decompositions
of $\rho\bar\rho\rho\bar\rho\cdots\bar\rho$ give objects of
a \textsl{tensor category}, where the morphisms are the
intertwiners of endomorphisms
and the tensor product operation is composition of endomorphisms.  
If we have finitely many irreducible endomorphisms
arising in this way, which is equivalent to the finite depth
condition, our tensor category is a \textsl{fusion category}, 
where we have the dual object for each object and we have
only finitely many irreducible objects up to isomorphisms.  
The higher relative commutants $M'\cap M_k$ are
described as intertwiner spaces like
$\mathrm{End}(\rho\bar\rho\rho\bar\rho\cdots\bar\rho)$ or
$\mathrm{End}(\rho\bar\rho\rho\bar\rho\cdots\rho)$.

In our setting, for a factor $M$, 
we have the \textsl{standard representation}
of $M$ on the Hilbert space $L^2(M)$, the  completion
of $M$ with respect to a certain inner product, and this
$L^2(M)$ also has a right multiplication by $M$ based on
Tomita-Takesaki theory.
For an endomorphism $\rho$ of $M$, we have a new $M$-$M$
bimodule structure on $L^2(M)$ by twisting the right action
of $M$ by $\rho$.  In this setting, all $M$-$M$ bimodules arise in 
this way, and we have a description of the above tensor
category in terms of bimodules.  Here the tensor product
operation is given by a \textsl{relative tensor product}
of bimodules over $M$.  For type II$_1$ factors,
we need to use this bimodule description to obtain the
correct tensor category structures.  
It is more natural to use type II$_1$ factors
in  statistical mechanics, and it is more natural to use
type III$_1$ factors in quantum field theory, but they give rise to 
 equivalent tensor categories, so if we are interested in
tensor category structure, including braiding, this difference
between type II$_1$ and type III$_1$ is not important.

\section{Subfactors and conformal field theory}

A two-dimensional conformal field theory is a particular example 
of a quantum field theory, but it is a rich source of deep
interactions with subfactor theory, so we treat this in an independent
section.

We start with the $(1+1)$-dimensional Minkowski space and consider
quantum field theory with \textsl{conformal symmetry}.  We
restrict a quantum field theory onto two light rays $x=\pm t$ and
compactify a light ray by adding a point at infinity.  The
resulting $S^1$ is our ``spacetime'' now, though space and time
are mixed into one dimension, and our symmetry group for
$S^1$ is now $\mathrm{Diff}(S^1)$, the orientation preserving
\textsl{diffeomorphism group} of $S^1$.  Our spacetime region is now
an \textsl{interval} $I$, a non-empty, non-dense open connected subset
of $S^1$.  For each such an interval $I$, we have a corresponding
von Neumann algebra $A(I)$ acting on a Hilbert space $H$ of states
containing the
vacuum vector.  \textsl{Isotony} means that we have $A(I_1)\subset A(I_2)$
if we have $I_1\subset I_2$.  \textsl{Locality} now means that
$[A(I_1),A(I_2)]=0$, if $I_1 \cap I_2=\O$.  Note that
spacelike separation gives this very simple disjointness.
Our spacetime symmetry group now is $\mathrm{Diff}(S^1)$,
and we have a
projective unitary representation $U$ on $H$.  
Conformal covariance asks for
$U_g A(I) U_g^*=A(gI)$ for $g\in \mathrm{Diff}(S^1)$.  \textsl{Positivity
of the energy} means that the
restriction of $U$ to the subgroup of rotations of $S^1$ gives
a one-parameter unitary group and its generator is positive.
In this setting, a family $\{A(I)\}$ of von Neumann algebras 
satisfying these axioms is called a \textsl{conformal net}.

A representation theory of a conformal net in the style of
Doplicher-Haag-Roberts now gives a \textsl{braiding} due to the 
low-dimensionality of the ``spacetime'' $S^1$.  This is a
certain form of the non-trivial commutativity of endomorphisms
up to inner automorphisms.  That is,
two representations give two endomorphisms $\lambda,\mu$
of a single von Neumann algebra $A(I_0)$ for some fixed interval
$I_0$, and we have a unitary $\varepsilon(\lambda,\mu)\in A(I)$
satisfying $\mathrm{Ad}(\varepsilon(\lambda,\mu))\lambda\mu=\mu\lambda$.
This unitary $\varepsilon(\lambda,\mu)$, sometimes called a
\textsl{statistics operator}, arises from the monodromy of moving
an interval in $S^1$ to a disjoint one and back, and satisfies various compatibility
conditions such as \textsl{braiding-fusion equations} for 
intertwiners as in \cite{FRS1, FG, L2}.  
Switching two tensor components corresponds to switching two
wires of a braid.  For two wires, we have an overcrossing
and an undercrossing.  They correspond to
$\varepsilon(\lambda,\mu)$ and $\varepsilon(\mu,\lambda)^*$.
In particular, if we fix an irreducible endomorphism and use it
for both $\lambda$ and $\mu$, we have a unitary representation
of the \textsl{braid group} $B_n$ for every $n$.
In the case of a higher-dimensional
Minkowski space, $\varepsilon(\lambda,\mu)$ gives a so-called
\textsl{degenerate} braiding, like the case of a group representation
where we easily have unitary equivalence of
$\pi\otimes\sigma$ and $\sigma\otimes\pi$ for two representations
$\pi$ and $\sigma$, but we now have a braiding in a more
non-trivial way on $S^1$.  It was proved  by Kawahigashi-Longo-M\"uger in \cite{KLM}
that if we have a certain finiteness of the representation theory of
a conformal net, called \textsl{complete rationality}, then
the braiding of its representation category is \textsl{non-degenerate}, and
hence it gives rise to a \textsl{modular tensor category} by definition.
A modular tensor category is also expected to be useful for
\textsl{topological quantum computations} as in  the work of Freedman-Kitaev-Larsen-Wang \cite{FKLW}.  This is
a hot topic in quantum information theory and many researchers work on
topological quantum information using the Jones polynomial and
its various generalisations.

It is a highly non-trivial task to construct examples of conformal nets.
The first such attempt started in a joint project of Vaughan
and Wassermann trying to construct a subfactor from a \textsl{positive
energy representation} of a loop group.
Wassermann \cite{Wa} then constructed conformal nets arising from
positive energy representations of the loop groups of $SU(N)$
corresponding to the Wess-Zumino-Witten models $SU(N)_k$, where
$k$ is a positive integer called a \textsl{level}.
These examples satisfy complete rationality as shown  by Xu in \cite{X4}.  
The conformal nets corresponding to $SU(2)_k$ give unitary representations
of the braid groups $B_n$ which are the same as the one given by Vaughan
from the Jones projections $e_j$.
Wassermann's construction has been generalised to other Lie groups by
Loke, Toledano Laredo and Verrill   in  dissertations supervised
by him, \cite{Loke, Tol, Ve},  see also \cite{Wa1}.
 Loke worked with projective unitary representations
of $\mathrm{Diff}(S^1)$ and obtained the \textsl{Virasoro nets}.

A relative version $\{A(I)\subset B(I)\}$ of a conformal net
for intervals
$I\subset S^1$  called a \textsl{net of subfactors} has been given in \cite{LR}.
Suppose that $\{A(I)\}$ is completely rational.  Assuming that
we know the representation category of $\{A(I)\}$, we would like to
know that of $\{B(I)\}$.  The situation is
similar to a group inclusion $H\subset G$ where we know representation
theory of $H$ and would like to know that of $G$.  In the group
representation case for $H\subset G$, we have  a \textsl{restriction} of 
a representation of $G$ to $H$ and an \textsl{induction} of a representation
of $H$ to $G$.  In the case of a net of subfactors, the restriction
of a representation of $\{B(I)\}$ to $\{A(I)\}$ is easy to define,
but the induction procedure is more subtle.  Our induction 
procedure is now called \textsl{$\alpha$-induction},  first
defined  by Longo-Rehren in \cite{LR} and studied 
by Xu \cite{X1},   B\"ockenhauer-Evans \cite{BE1, BE2, BE3, BE4},   and B\"ockenhauer-Evans-Kawahigashi  \cite{BEK1, BEK2},
also in connection to Ocneanu's graphical calculus on Coxeter-Dynkin diagrams
in the last two papers.
(In these two papers, this $\alpha$-induction
is studied in the more general context
of abstract modular tensor categories of endomorphisms rather than
conformal field theory.  For an $A$-$A$  bimodule $X$,  then  the tensor product $X \otimes B$ can be regarded as a $B$-$B$ module if one uses the braiding to let $B$ act on the left.)
Take a representation of $\lambda$ of $\{A(I)\}$ which is
given as an endomorphism of $A(I_0)$ for some fixed interval $I_0$.
Then using the braiding on the representation category of
$\{A(I)\}$, we define an endomorphism $\alpha_\lambda^\pm$ of
$B(I_0)$ where $\pm$ represents a choice of a positive or negative
braiding, $\varepsilon^\pm(\lambda,\theta)$, where $\theta$ represents
the dual canonical endomorphism of the subfactor $A(I)\subset B(I)$.
This nearly gives a representation of
$\{B(I)\}$, but not exactly.  It turns out that
the irreducible endomorphisms 
arising both from a positive induction and a negative one
exactly correspond to those arising from irreducible
representations of $\{B(I)\}$.
The braiding of the representation category of $\{A(I)\}$
gives a finite dimensional unitary representation of
$SL(2,\mathbb{Z})$ through the so-called \textsl{$S$- and $T$-matrices}.
 B\"ockenhauer-Evans-Kawahigashi \cite{BEK1}  showed that the matrix
$Z_{\lambda,\mu}=\langle \alpha^+_\lambda, \alpha^-_\mu\rangle$,
where $\lambda,\mu$ label irreducible representations of
$\{A(I)\}$ and the symbol $\langle\cdot,\cdot\rangle$ counts
the number of common irreducible endomorphisms including 
multiplicities, satisfies the following properties:

\begin{enumerate}
\item We have $Z_{\lambda,\mu}\in \{0,1,2,\dots\}$.
\item We have $Z_{0,0}=1$, where the label $0$ denotes the
vacuum representation.
\item The matrix $Z$ commutes with the image of the
representation of $SL(2,\mathbb{Z})$.
\end{enumerate}

Such a matrix $Z$ is called a \textsl{modular invariant}, because
$PSL(2,\mathbb{Z})$ is called the \textsl{modular group}.
For a given completely rational
conformal net (or more generally, a given modular
tensor category), we have only finitely many modular invariants.
Modular invariants naturally appear as \textsl{partition functions}
in 2-dimensional conformal field theory and they have been
classified for several concrete examples since  Cappelli-Itzykson-Zuber \cite{CIZ} for
the $SU(2)_k$ models and the \textsl{Virasoro nets} with $c<1$, where
$c$ is a numerical invariant called the \textsl{central charge}.
It takes a positive real value, and if $c<1$, then it is of the form
$1-6/m(m+1)$, $m=3,4,5,\dots$ by  Friedan-Qiu-Shenker \cite{FQS}  and Goddard-Kent-Olive \cite{GKO}.
This number arises from a projective unitary representation of
$\mathrm{Diff}(S^1)$ and its corresponding unitary representation of
the \textsl{Virasoro algebra}, a central extension of the complexification 
of the Lie algebra arising from $\mathrm{Diff}(S^1)$.
Note that some modular invariants defined by the above three
properties do not necessarily correspond to \textsl{physical}
ones arising as partition functions in conformal field theory.  Modular invariants
arising from  $\alpha$-induction are physical in 
this sense.

 The action of the $A$-$A$ system 
on the $A$-$B$  sectors (obtained by decomposing
$\{ \iota \lambda =\alpha^{\pm}_{\lambda} \iota : \lambda  \in$ $A$-$A\}$
into irreducibles where $\iota : A \rightarrow B$ is the inclusion)
 gives naturally a representation of the fusion
rules of the Verlinde ring:
$G_{\lambda}G_{\mu} = \sum N_{\lambda \mu}^{\nu} G_{\nu}\,,$
 with matrices $ G_{\lambda} = [G_{\lambda a}^{b} : a, b \in$ $A$-$B$ sectors].
Consequently, the matrices $G_{\lambda}$ will be described by the
 same eigenvalues but with possibly different multiplicities.
 B\"ockenhauer-Evans-Kawahigashi \cite{BEK2}  showed that these   multiplicities are given exactly by the
 diagonal part of the modular invariant:
${\rm spectrum} (G_{\lambda}) = \{S_{\lambda\kappa}/S_{0\kappa}  
: {\rm with \,multiplicity}\, Z_{\kappa\kappa} \}\,.
$
This is called a {\it nimrep}  -- a non-negative integer matrix representation.
Thus a  physical modular invariant is automatically
 equipped with a  compatible nimrep whose spectrum is described by the diagonal part of
the modular invariant.  The case of
$SU(2)$ is just the $A$-$D$-$E$ classification of Cappelli-Itzykson-Zuber \cite{CIZ}
with the $A$-$B$ system yielding the associated (unextended) Coxeter-Dynkin graph. Since 
there is an $A$-$D$-$E$ classification of  matrices of norm less than $2$, we can recover independently of Cappelli-Itzykson-Zuber \cite{CIZ} that there are unique modular invariants corresponding to the three exceptional $E$ graphs.

If we use only positive $\alpha$-inductions for a given modular
tensor category, we still have a fusion category of endomorphisms,
but no braiding in general.  This is an
example of a \textsl{module category}.  For the tensor category $Rep(G)$ of representations of a finite group $G$, all module categories are of the form $Rep(H, \chi)$ for the projective representations 
with $2$-cocycle $\chi$ for a subgroup $H$  \cite{O2}.  For this reason, module categories have also been called {\it quantum subgroups}.  Such categories
have been studied in a more general  categorical  context by Ostrik in \cite{Os}.
 However, Carpi, Gaudio, Giorgetti and  Hillier  \cite{CGGH}, have shown that for unitary fusion categories, such as those that occur in subfactor theory  or arise from loop groups, that all module categories are equivalent to unitary ones.
For the conformal nets corresponding to $SU(2)_k$, the
module categories or  quantum subgroups are labeled with all the Coxeter-Dynkin diagrams $A_n$,  $D_n$
and $E_{6,7,8}$.  Here there is a coincidence with the affine $A$-$D$-$E$ classification of finite subgroups of 
$SU(2)$. Di Francesco and Zuber \cite{DZ} were motivated to try to relate $SU(3)$ modular invariants with subgroups of $SU(3)$.
There is a partial match but this is not helpful. In general whilst the number of finite subgroups of $SU(n)$ grows with $n$, the number of exceptional modular invariants, beyond the obvious infinite series, does not.

If we have a net of subfactors $\{A(I)\subset B(I)\}$ 
with $\{A(I)\}$ being a completely rational conformal net,
then the restriction of the vacuum representation of $\{B(I)\}$
to $\{A(I)\}$ gives a \textsl{local $Q$-system} in the sense of
 Longo
\cite{L3}.  This notion is essentially the same as a 
\textsl{commutative Frobenius algebra}, a special case of
an \textsl{algebra in a tensor category}, in the algebraic or categorical 
literature.  This $Q$-system is a triple consisting of an
object and two intertwiners.  Roughly speaking, the object
gives $B(I)$ as an $A(I)$-$A(I)$ bimodule and the intertwiners
give the multiplicative structure on $B(I)$.
Our general theory of $\alpha$-induction shows that the corresponding 
modular invariant $Z$ for the modular tensor category of
representations of $\{A(I)\}$ recovers this object.
Since we have only finitely
many modular invariants for a given modular tensor category,
we have only finitely many  objects for a local $Q$-system.
It is known that each object has only finitely many
local $Q$-system structures, and we thus have only finitely many
local $Q$-systems, which means that we have only finitely many
possibilities for extensions $\{B(I)\}$ for a given $\{A(I)\}$.

For some concrete examples of $\{A(I)\}$, we can classify all
possible extensions.  In the case of the $SU(2)_k$ nets,
such extensions were studied in the context of $\alpha$-induction
in \cite{BEK1}  by B\"ockenhauer-Evans-Kawahigashi  and it was shown in \cite{KL1}   by Kawahigashi-Longo
that they exhaust all possible extensions.  
(A similar classification based on quantum groups was first
given in \cite{KO}.)
They correspond to the Coxeter-Dynkin diagrams $A_n$, $D_{2n}$, $E_6$ and
$E_8$.  The $A_n$ cases are the $SU(2)_k$ nets themselves,
the $D_{2n}$ cases are given by
\textsl{simple current extensions} of order 2, 
and the $E_6$ and $E_8$ cases are given by
\textsl{conformal embeddings} $SU(2)_{10}\subset SO(5)_1$
and $SU(2)_{28}\subset (G_2)_1$, respectively.
These correspond to
\textsl{type I extensions} in Itzykson-Zuber \cite{CIZ}, 
B\"ockenhauer-Evans \cite{BE4}.
Type II extensions corresponding to $D_{2n+1}$ and
$E_7$ arise from extensions of the $SU(2)_k$ nets \textsl{without} locality.
In general \cite{BE4} for a physical modular invariant $Z$ there are  by B\"ockenhauer-Evans   local chiral extensions $N(I) \subset M_+(I)$ and $N(I) \subset M_-(I)$ 
with local $Q$-systems naturally associated to the vacuum column $\{Z_{\lambda, 0}\}$  and vacuum row $\{Z_{0, \lambda}\}$ respectively. These extensions are indeed maximal and should
be regarded as the subfactor version of left- and right maximal extensions of the chiral algebra. The representation theories or modular tensor categories
of $M_\pm$  are then  identified. For example, the  $E_7$ conformal net or module category is a then a twist or auto-equivalence on the left and right local $D_{10}$ extensions which form the type I parents.
This reduces the analysis to understanding first local extensions  and then classifying auto-equivalences to identify the  two left and right local extensions.  For $SU(2)$ there are only three exceptional modular invariants $E_{6,7,8}$, and in general one expects,  e.g. \cite{O1},  for a WZW model that there are only a finite number of exceptionals  beyond the infinite series of the trivial, orbifolds and their conjugates. Schopieray \cite{S} using $\alpha$-induction found bounds for levels of exceptional invariants for rank $2$ Lie groups, and Gannon \cite{G} extended this for higher rank with improved lower bounds using Galois transformations as a further tool.  Edie-Michell has undertaken extensive studies of  auto-equivalences \cite{EM}.  The realisation by Evans-Pugh \cite{EvPu} of $SU(3)$-modular invariants as full CFT's, announced in \cite{O}, is based on the classification of Gannon \cite{G1} of $SU(3)$ modular invariants, and the classification by Evans-Pugh of full $SO(3)$ theories or $SO(3)$ module categories is in \cite{EvPu1}.

For a general conformal net, we always have a subnet generated
by the projective unitary representation of $\mathrm{Diff}(S^1)$, 
which is called the \textsl{Virasoro net}, so a conformal net is always
an extension of the Virasoro net. Through a unitary representation
of the Virasoro algebra, a conformal net has a numerical
invariant $c$, the central charge.  
The Virasoro net is completely rational if $c<1$, so the above
classification scheme applies to this case, and we have
a complete classification of conformal nets with $c<1$   by Kawahigashi-Longo  in
\cite{KL1}, where they are shown to be in a bijective correspondence with the 
type I modular invariants  of Cappelli-Itzykson-Zuber in \cite{CIZ}.  Four of exceptional
modular invariants involving the Dynkin diagrams $E_6$ and $E_8$
give exceptional conformal nets.  Three of them are given by 
the \textsl{coset construction}, but the other one gives a new
example.  Similarity between
discreteness of the Jones index values below $4$ and 
discreteness of the central charge value below $1$ has
been pointed out since the early days of subfactor theory \cite{J6},
and we have  an $A$-$D$-$E$ classification of subfactors
with index below $4$ as in Popa \cite{P1} (also
see \cite{O, EK})  and an $A$-$D$-$E$ classification of the 
modular invariants of the Virasoro minimal models of  Capelli-Itzykson-Zuber
\cite{CIZ}.  We then have natural understanding of
classification of conformal nets with $c<1$ in this context.

$K$-theory has had a role in relating subfactor theory with statistical mechanics and conformal field theory.  The phase transition in the two dimensional Ising model is    analysed through an analysis of the ground states of the one dimensional quantum system arising 
from the transfer matrices. This is manifested by Araki-Evans through
a jump in the Atiyah-Singer  mod-$2$ index of Fredholm operators \cite{AE}. Here Kramers-Wannier high-temperature duality is effected by the shift endomorphism $\rho$ on the corresponding Jones projections $e_j \rightarrow e_{j+1}$
 which leads, Evans \cite{Eva}, to  the Ising fusion rules $\rho^2 = 1 + \sigma$, where $\sigma$ is the symmetry formed from interchanging $+$ and $-$ states, see also Evans-Gannon \cite{EG6}.
 The tensor category of the Verlinde ring of compact Lie groups, or doubles of finite groups has been described by Freed-Hopkins-Teleman \cite{FHT} through the  twisted equivariant $K$-theory of the group acting on itself by conjugation. This has allowed the interchange of ideas between the subfactor approach and a $K$-theory approach to conformal field theory, employing $\alpha$-induction and modular invariants as bi-variant Kasparov $KK$-elements   by Evans-Gannon
\cite{EG1, EG2,  EG6}. In a similar spirit, regarding $K$-theory in terms of projective modules, a finitely generated modular tensor category  can be realised  by Aaserud-Evans \cite{AsE} as $C^*$-Hilbert modules. This applies to  the modular tensor categories  of Temperley-Lieb-Jones associated to quantum $SU(2)$, or more generally those of loop groups  -- as well as quantum doubles such as that of  the Haagerup subfactor which we will focus on in the final section.
This also gives a framework for braided tensor categories acting on some $C^*$-algebras as a quantum symmetry.

\section{Vertex operator algebras}

We have another, more algebraic, 
mathematical axiomatisation for a chiral
conformal field theory, namely, a \textsl{vertex operator algebra}.
Since a conformal net and a vertex operator algebra are both
mathematical formulations of the same physical theory, they
naturally have close relations.  We now explain those here.

A quantum field on the ``spacetime'' $S^1$ is an operator-valued
distribution on $S^1$, so it has a Fourier expansion with 
operator coefficients.  In this axiomatisation,
we have a $\mathbb{C}$-vector space $V$ which
is a space of finite energy vectors and is supposed to give 
the Hilbert space of  states after completion.
For each vector $u\in V$, we have a formal series
$Y(u,z)=\sum_{n\in\mathbb{Z}} u_n z^{-n-1}$ with a
formal variable $z$ and linear operators $u_n$ on $V$, which
corresponds to the Fourier expansion of a quantum field
acting on the completion of $V$.  This correspondence
from a vector to a formal series is called the
\textsl{state-field correspondence}.  We have two
distinguished vectors, the \textsl{vacuum vector} and 
the \textsl{Virasoro vector}.
 The Fourier coefficients
of the latter give the Virasoro algebra.  The locality axiom in this
setting says that for $u,v\in V$, we have a sufficiently large
positive integer $N$ satisfying
$(z-w)^N [Y(u,z), Y(v,w)]=0$.  Roughly speaking, this means
$Y(u,z)Y(v,w)=Y(v,w)Y(u,z)$ for $z\neq w$.

The origin of this notion of a vertex operator algebra is as follows.
A classical \textsl{elliptic modular function} 
\[
j(\tau)=1728\frac{g_2(\tau)^3}{g_2(\tau)^3-27 g_3(\tau)^2},
\]
where $\mathrm{Im}\;\tau>0$ and
$g_2(\tau)$ and $g_3(\tau)$ are defined by the 
\textsl{Eisenstein series}, has the
following Fourier expansion with $q=\exp(2\pi i\tau)$.
\[
j(\tau)=q^{-1}+744+196884q+21493760q^2+\cdots.
\]
McKay noticed that the coefficient $196884$ is very close
to $196883$, which is the dimension of the lowest-dimensional
non-trivial irreducible representation of the \textsl{Monster group}.
Recall that the Monster group is the largest among the 26
sporadic finite simple groups in terms of its order which is around $8\times 10^{53}$.  It turns out that
we have a similar relation 
$21493760=1+196883+21296876$, where $21296876$ is the
dimension of the next lowest-dimensional irreducible
representation of the Monster group.  Based on this and
many other pieces of information on modular functions,
Conway-Norton \cite{CN} made the following \textsl{Moonshine conjecture}.

\begin{enumerate}
\item There is a graded $\mathbb{C}$-vector space 
$V=\bigoplus_{n=0}^\infty V_n$ with \textsl{some algebraic structure}
whose automorphism group isomorphic to the Monster group.
\item For any element $g$ in the Monster group,
$\sum_{n=0}^\infty \mathrm{Tr}(g|_{V_n}) q^{n-1}$ is the 
\textsl{Hauptmodul} for a genus 0 subgroup of $SL(2,\mathbb{R})$,
where $g|_{V_n}$ is the  linear action of an automorphism $g$ on $V_n$.
\end{enumerate}

Frenkel-Lepowsky-Meurman \cite{FLM} gave a precise definition
of a  \textsl{certain algebraic structure} as a vertex operator algebra,
constructed the \textsl{Moonshine vertex operator algebra} $V^\natural$,
and proved that its automorphism group is exactly the Monster group.
They first constructed a vertex operator algebra from the \textsl{Leech
lattice}, an exceptional $24$-dimensional lattice giving
the densest sphere packing in dimension 24, and applied
the \textsl{twisted orbifold construction} for the order two
automorphism of the vertex operator algebra arising from the
multiplication by $-1$ on the Leech lattice to obtain $V^\natural$.
Borcherds \cite{B} next proved the remaining part of the Moonshine
conjecture.  The construction of a vertex operator algebra from
an even lattice has an operator algebraic counterpart for 
a conformal net given in  Dong-Xu  \cite{DX}.  The operator algebraic
counterpart of the Moonshine vertex operator algebra has been
constructed as the \textsl{Moonshine net} in  Kawahigashi-Longo \cite{KL2}.
Frenkel-Zhu gave a construction of vertex operator algebras
from affine Kac-Moody and Virasoro algebras in \cite{FZ},
and this corresponds to the construction of conformal nets
of Wassermann \cite{Wa}, Loke, Toledano Laredo and Verrill.

We have constructions of new examples of vertex operator algebras
or conformal nets from known ones as follows.  
\begin{enumerate}
\item A tensor product
\item Coset construction
\item Orbifold construction
\item An extension using a $Q$-system
\end{enumerate}
In the operator algebraic setting, the \textsl{coset construction}
gives a relative commutant $A(I)'\cap B(I)$ for an
inclusion $\{A(I)\subset B(I)\}$ of conformal nets
of infinite index.  The \textsl{orbifold construction} gives 
a fixed point conformal subnet given by an automorphic
action of a finite group.  These constructions
for conformal nets have been studied  by Xu  in  
\cite{X2} and \cite{X3}, respectively. The extension of
a local conformal net using a $Q$-system was
first studied   by Kawahigashi-Longo in \cite{KL1} for constructing exceptional
conformal nets, and this was extended   by Xu in \cite{X5}.  The vertex operator
algebra counterpart has been studied   by Huang-Kirillov-Lepowsky  in \cite{HKL}.
Xu has shown that various subfactor techniques are quite powerful
even for purely algebraic problems in vertex operator algebras.

From the above results, it is clear that we have close
connections between conformal nets and vertex operator
algebras, as expected, but it is more desirable to have a direct
construction of one from the other.  The relation between
the two should be like the one between Lie groups and Lie
algebras, and the former should be given by ``exponentiating''
the latter.  Such a construction
was first given in   Carpi-Kawahigashi-Longo-Weiner \cite{CKLW}.  That is, we have a
construction of a conformal net from a vertex
operator algebra with \textsl{strong locality}
and we also recover the original vertex operator algebra
from this conformal net.  (Note that we obviously need
\textsl{unitarity} for a vertex operator algebra for such construction,
since we need a nice positive definite inner product on $V$.
This unitarity is a part of the strong locality assumption.
There are many vertex operator algebras without unitarity,
and they may be related to operator algebras through different
routes such as planar algebras.)
In addition to an abstract definition of strong locality,
concrete sufficient conditions for this have been also given
in \cite{CKLW}.  This correspondence
between vertex operator algebras and conformal nets has
been vastly generalised recently in
  Gui  \cite{G},  Raymond-Tanimoto-Tener  \cite{RTT} and  Tener  \cite{T1, T2, T3}
including identification of their representation categories,
and this is a highly active area of research today.  Some of
them started from dissertations supervised by Vaughan.

\section{Other directions in conformal field theory}

The classification of subfactors with index less than $4$ has
an $A$-$D$-$E$ pattern.  That is, the flat connections
given in Fig.\ref{Dynkin} give a complete list of hyperfinite
II$_1$ subfactors with index less than $4$;  see the review  \cite{JMS}.
It naturally has connections to
many other topics in mathematics and physics where
$A$-$D$-$E$ patterns appear.  At the index
value equal to $4$, we still have a similar $A$-$D$-$E$ 
classification based on extended Dynkin diagrams due to Popa \cite{P1}.  They
correspond to subgroups of $SU(2)$, and the extended
Coxeter-Dynkin diagrams appear through the \textsl{McKay correspondence}.
These subfactors arise as simultaneous fixed point algebras
of actions of a  subgroup of $SU(2)$ on 
\[
\mathbb{C}\otimes M_2(\mathbb{C})\otimes M_2(\mathbb{C})\otimes\cdots
\subset M_2(\mathbb{C})\otimes M_2(\mathbb{C})
\otimes M_2(\mathbb{C})\otimes\cdots
\]
with infinite tensor products
of the adjoint actions, possibly with extra cohomological twists  as in the classification of
periodic actions by Connes \cite{Co1} and for finite group actions by Vaughan in his thesis \cite{J} on the hyperfinite II$_1$ factor.
The subfactors with index less than
4 can be  regarded as ``quantum'' versions of this construction.

We have a quite different story above the index value $4$.
Haagerup searched for subfactors of finite depth above
index value $4$ and found several candidates of 
the principal graphs.  The smallest index value among them
is $(5+\sqrt{13})/2$ and he proved this index value is indeed
attained by a subfactor, which is now called the \textsl{Haagerup
subfactor} \cite{AH}.  A similar method also produced the
\textsl{Asaeda-Haagerup subfactor} in \cite{AH}.
New constructions of the Haagerup
subfactor were given in  Izumi  \cite{I2} and in   Peters  \cite{Pe}.  The latter is
based on the planar algebra machinery, and has been extended to
the construction of the \textsl{extended Haagerup subfactor}.
Today we have a complete classification of subfactors of finite
depth with index
value between 4 and 5 as reviewed in \cite{JMS}, and we have
five such subfactors (after identifying $N\subset M$ and $M\subset M_1$):
the Haagerup subfactor, the Asaeda-Haagerup subfactor, the
extended Haagerup subfactor, the Goodman-de la Harpe-Jones \cite{GHJ}
subfactor and the Izumi-Xu subfactor.  The latter two are now
understood as  arising from conformal embeddings  $SU(2) \rightarrow E_6$ and $G_2 \rightarrow E_6$.
 If a  subfactor arises from a connection on a finite graph $\Gamma$,  it may not have principal graph or standard invariant based on $\Gamma$ as happens with  $D_{2n+1}$ or $E_7$. Any graph whose norm squared is  in the range $(4,5)$ but  is not one of the five allowed values  can only have $A_\infty$ as principal graph just like what happens with $E_{10}$ by  unpublished work of Ocneanu, Haagerup and Schou.

The fusion categories arising from the Haagerup
subfactor do not have a braiding, but their \textsl{Drinfel$'$d
center} always gives a modular tensor category.  Izumi gave a
new construction of the Haagerup subfactor and computed
the $S$- and $T$-matrices of its Drinfel$'$d center in
\cite{I1, I2}, using endomorphisms of the Cuntz algebra.
It is an important problem whether
an arbitrary modular tensor category is realised as the 
representation category of a conformal net or not, and this
particular case of the Drinfel$'$d center of the fusion category of 
the Haagerup has caught much attention.  Note that all the known
constructions \cite{AH, I2, Pe} of the Haagerup subfactor are based on
algebraic or combinatorial computations. There is little conceptual understanding of this subfactor and its double and it is not clear
at all whether they are related to statistical mechanics or
conformal field theory.  Evidence
in the positive direction has been given   by Evans-Gannon in \cite{EG, EG6}.
 They found characters
for the representation of the modular group $SL(2, \mathbb Z)$ arising from the braiding and 
showing that this modular data, their $S$ and $T$ matrices and fusion rules have a simple expression in terms of a grafting of the double of the dihedral group $S_3$ and $SO(13)_2$, or indeed the
orbifolds of two Potts models or quadratic (Tambara-Yamagami) systems based on $\mathbb Z_3 \times \mathbb Z_3$ and $\mathbb Z_{13}$ respectively.

 Information about a conformal field theory from the scaling limit of a statistical mechanical model may be detected from the underlying statistical mechanical system. 
Cardy \cite{Cardy} argued from conformal invariance for a critical statistical system,
that the central charge $c$ may be computed from the asymptotics of the
partition function and transfer matrices on a periodic rectangular  lattice.
This has been well studied for the ABF, Q-state Potts models for Q=2,3,4 and certain ice-type models; see  \cite[pages 453--454]{EK}. In this spirit, numerical computations have been made using transfer matrices built from associator or certain $6j$ symbols for a Haagerup system, though not the double.  These give a value of $c=2$ (or around  $2$) \cite{HLOTT, VLVDWOHV, LZR}. 
However   the results shown there  and these methods do not show that if there is a CFT at  $c=2$ (or around $2$)  that it is not a known one and that if there is a CFT that its representation theory is related to  the representation theory of the double of the Haagerup. Recall  what we described in a preceding paragraph, that a subfactor constructed from a graph may not reproduce the graph through its invariants.

The first non-trivial reconstructions of conformal field theories were achieved   by Evans-Gannon for the twisted doubles of finite groups and the orbifolds of Potts models
\cite{EG5, EG6}.  
Whilst von Neumann algebras and subfactors are inherently unitary, non unitary theories have been analysed   by Evans-Gannon from ideas derived from subfactors.
 This includes the Leavitt path algebras to replace Cuntz algebras in constructing non-unitary tensor categories of algebra endomorphisms which do not necessarily preserve the $*$-operation \cite{EG4}.
 These and non-unitary planar algebras could also be a vehicle to understand non-semisimple and logarithmic conformal field theories.

In attempting to  construct  a conformal net realizing a
given modular tensor category, a natural idea is to 
construct algebras as certain limit through finite dimensional
approximations.  We then use lattice approximation of the
circle $S^1$, but diffeomorphism symmetry is lost in this
finite dimensional approximation, so it is a major problem how to recover
diffeomorphism symmetry.  Vaughan studied this problem,
used \textsl{Thompson's groups} as approximations of $\mathrm{Diff}(S^1)$,
and obtained various interesting representations of Richard Thompson's
groups \cite{BJ1, J8, J9}.  Though he proved in \cite{J9}
that translation operators arising as a limit of translations 
for the $n$-chains do not extend to a translation group that is
strongly continuous at the origin, these representations are
interesting in their own right. 
The clarity of his formalism and analysis, led to concise and elegant
proofs of the previously difficult facts that the Thompson groups did not have Kazhdan's property T and with his Berkeley student Arnaud Brothier \cite{BJ2} that 
the Thompson's group T did not have the Haagerup property. New results also followed -- certain wreath products of groups have the Haagerup property
by taking the group of fractions of group labelled forests. Taking a functor from binary forests to Conway tangles,
replacing a fork by an elementary tangle, Vaughan could show that every link arises in this way from the fraction of a pair of forests
just as braids yield all links through taking their closures ---
providing another unexpected bridge with knots and links  \cite{J11}.
He further studied related
problems on scale invariance of transfer matrices on \textsl{quantum
spin chains}, introduced
two notions of scale invariance and weak scale invariance,
and gave conditions for transfer matrices and nearest
neighbour Hamiltonians to be scale invariant or weakly scale
invariant \cite{J10}.\\

\begin{footnotesize}
\noindent{\it Acknowledgement.}
We wish to thank Sorin Popa for discussions and comments on the manuscript and are grateful to George Elliott and  Andrew Schopieray for extremely careful proofreading.
\end{footnotesize}
\bibliographystyle{amsplain}

\end{document}